\documentclass{raa_twocolumn}

\usepackage{graphicx,times}             
\usepackage{natbib}
\usepackage{amssymb,amsmath}
\usepackage{float}
\bibpunct{(}{)}{;}{a}{}{,}

\newcommand{\ptz}{photo-\textit{z}}	
\newcommand{\spz}{spec-\textit{z}}

\usepackage[pagebackref=true]{hyperref}

\begin{document}

  \title{Improving Photometric Redshift Estimation for CSST Mock Catalog Using SED Templates Calibrated with Perturbation Algorithm}

   \volnopage{Vol.0 (2025) No.0, 000--000}      
   \setcounter{page}{1}          

    \author{Yicheng Li 
        \inst{1}
    \and Liping Fu${^*}$
        \inst{1,2}
    \and Zhu Chen${^*}$
        \inst{1}
    \and Zhijian Luo${^*}$
        \inst{1}   
    \and Wei Du
        \inst{1}
    \and Yan Gong
        \inst{3,4}
    \and Xianmin Meng
        \inst{4}
    \and Junhao Lu
        \inst{1}
    \and Zhirui Tang
        \inst{1}
    \and Pengfei Chen
        \inst{1}
    \and Shaohua Zhang
        \inst{1}
    \and Chenggang Shu
        \inst{1}
    \and Xingchen Zhou
        \inst{3}
    \and Zuhui Fan
        \inst{5}
   }

    \institute{
        Shanghai Key Lab for Astrophysics, Shanghai Normal University, Shanghai 200234, China; { fuliping@shnu.edu.cn, zhuchen@shnu.edu.cn, zjluo@shnu.edu.cn}\\
        \and
        Center for Astronomy and Space Sciences, China Three Gorges University, Yichang 443000, People’s Republic of China \\
        \and
        Key Laboratory of Space Astronomy and Technology, National Astronomical Observatories, Chinese Academy of Sciences, 20A Datun Road, Beijing 100101, China\\
        \and
        Science Center for China Space Station Telescope, National Astronomical Observatories, Chinese Academy of Sciences, 20A Datun Road, Beijing 100101, China\\
        \and
        South-Western Institute for Astronomy Research, Yunnan University, Kunming 650500, China\\
\vs\no
    {\small Received 20xx month day; accepted 20xx month day}}

\abstract{
Photometric redshifts of galaxies obtained by multi-wavelength data are widely used in photometric surveys because of its high efficiency. Although various methods have been developed, template fitting is still adopted as one of the most popular approaches. Its accuracy strongly depends on the quality of the Spectral Energy Distribution (SED) templates, which can be calibrated using broadband photometric data from galaxies with known spectroscopic redshifts. Such calibration is expected to improve photometric redshift accuracy, as the calibrated templates will align with observed photometric data more closely.
The upcoming China Space Station Survey Telescope (CSST) is one of the Stage IV surveys, which aiming for high precision cosmological studies. To improve the accuracy of photometric redshift estimation for CSST, we calibrated the CWW+KIN templates using a perturbation algorithm with broadband photometric data from the CSST mock catalog. This calibration used a training set consisting of approximately 4,500 galaxies, which is 10\% of the total galaxy sample. The outlier fraction and scatter of the photometric redshifts derived from the calibrated templates are 2.55\% and 0.036, respectively. Compared to the CWW+KIN templates, these values are reduced by 34\% and 23\%, respectively. 
This demonstrates that SED templates calibrated with a small training set can effectively optimize photometric redshift accuracy for future large-scale surveys like CSST, especially with limited spectral training data.
}
   \authorrunning{Li et al.}            
   \titlerunning{SED Templates Calibration for CSST Mock Catalog}  

   \maketitle

%
%
\section{Introduction}           
\label{sect:intro}

Estimating the redshifts of astronomical objects is essential for advancing extragalactic and cosmological studies (see \citealt{Salvato2019, Newman2022} for a review). Redshifts can be determined using two primary methods: spectroscopic and photometric observations. Spectroscopic redshifts (\spz) are obtained by measuring wavelength shifts in spectral line features. While this method provides high accuracy, obtaining \spz~for faint, distant objects is challenging and time consuming (\cite{Newman2015}). In contrast, photometric redshifts (\ptz) are estimated using multi-band photometric data, which capture broad-band features such as the Lyman and Balmer breaks. Although \ptz~sacrifices some accuracy compared to spectroscopic methods due to the limited resolution of photometric data, it is significantly more efficient, enabling redshift estimation from fewer exposures across all detected sources. This efficiency makes \ptz~a practical choice for many observational analyses.

There are two primary methods for deriving \ptz: machine learning and the spectral energy distribution (SED) fitting method. In principle, any regression based machine learning technique can be applied to estimate \ptz, including classical support vector machines (\citealt{Wadadekar2005}; \citealt{Jones2017}), decision tree based bagging methods such as random forests, or boosting algorithms (e.g., \citealt{Carliles2010}; \citealt{Dalmasso2020}; \citealt{Zhou2021}). Additionally, simple neural networks have been widely used (e.g., \citealt{Firth2003}; \citealt{Collister2004}; \citealt{Cavuoti2017}; \citealt{Razim2021}; \citealt{Zhou2021}). 

Machine learning techniques typically use magnitudes and colors as input features, but can also incorporate more digitized features correlated with redshift, such as morphological parameters, to improve estimation accuracy (\citealt{DIsanto2018}; \citealt{Gomes2018}). Furthermore, combining digitized features with photometric images to train hybrid deep neural networks (\citealt{Zhou2022}) or employing advanced architectures like Long Short Term Memory (LSTM) networks (\citealt{luo2024photometric}) has shown potential for achieving even higher precision. 

However, machine learning methods have inherent limitations. Their accuracy strongly depends on the quality and size of the training sample. Moreover, the lack of spectroscopic data for faint galaxies in the training set limits their ability to estimate \ptz~accurately through extrapolation.

The basic idea of SED fitting is to find the best match between observed multi-band photometric data and redshifted galaxy SED templates. These templates can be derived from observations (e.g., \citealt{Coleman1980}; \citealt{Kinney1996}; \citealt{Assef2010}) or from stellar population synthesis models (e.g., \citealt{Bruzual1993, Bruzual2003}; \citealt{Fioc1997}; \citealt{Maraston2005}). Many established SED fitting \ptz~estimators (e.g., HyperZ, \citealt{Bolzonella2000}; LePhare, \citealt{Arnouts1999}; BPZ, \citealt{Benitez2000}; EAZY, \citealt{Brammer2008}) can utilize either type of template set. These methods can provide redshift estimates as long as the redshifted template SEDs cover the relevant wavelength range of the observed filters. 

A key limitation of SED fitting is the representativeness of the template set for a given observational data set. Stellar population synthesis models involve numerous physical assumptions, and empirical templates often derived from local, bright galaxies may not capture the full range of galaxy properties across different epochs. As a result, evolutionary effects at higher redshifts can introduce additional uncertainties, reducing the accuracy of SED fitting approaches.

The current and next generation of large scale surveys, such as the Dark Energy Survey (DES, \citealt{2005astro.ph.10346T}), the Vera C. Rubin Observatory Legacy Survey of Space and Time (\citealt{Ivezic2008}; \citealt{Abell2009}), the Euclid Survey (\citealt{Laureijs2011}; \citealt{Euclid2024}), and the Roman Space Telescope Survey (\citealt{2012arXiv1208.4012G}; \citealt{Akeson2019}), will provide unprecedented amounts of photometric data. For many extragalactic objects in these datasets, obtaining \spz~will be impractical. To fully utilize these extensive datasets, it is crucial to improve and thoroughly assess the performance of \ptz~ methods when applied to these surveys (\citealt{Newman2022}).

The same considerations apply to the China Space Station Survey Telescope (CSST) (\citealt{Zhan2011}; \citealt{Zhan2018}; \citealt{Cao2018}). The CSST will employ seven photometric bands, ranging from near-ultraviolet (NUV) to near-infrared (NIR), namely NUV, $u$, $g$, $r$, $i$, $z$ and $y$, that cover wavelengths of approximately 2,500\,\(\text{\AA}\) to 10,000\,\(\text{\AA}\). The 5$\sigma$ magnitude limits for these bands are 25.4, 25.4, 26.3, 26.0, 25.9, 25.2, and 24.4, respectively. During its 10-year mission, CSST will survey a total of 17,500\,deg$^2$ of the sky, performing photometric and spectroscopic observations simultaneously. One of its primary objectives is to test theoretical cosmological models through weak gravitational lensing, which heavily relies on accurate redshift estimations.

This paper is part of a series that evaluates the performance of various \ptz~techniques to meet the scientific goals of the CSST using a mock CSST catalog. \cite{Lu2024rf} estimate \ptz s from CSST mock fluxes using a weighted random forest method. \cite{luo2024imputation} employ a deep learning method called generative adaptive imputation networks (GAIN) to impute missing photometric data, thus reducing the impact of data gaps on \ptz~estimation and improving precision. \cite{luo2024photometric} propose a new approach that relies solely on flux measurements from different observed filters to accurately predict CSST photo-$z$ by constructing a deep learning model based on Recurrent Neural Networks (RNN) with LSTM units.

In this work, we focus on improving the accuracy of \ptz~derived from the CSST mock catalog using SED fitting and calibrated SED templates (i.e., a template learning approach). Because the actual SEDs of galaxies may differ from the initial input templates, calibration of these templates is essential. Several studies have addressed template calibration and optimization (\citealt{benitez2004ck, Budavari2000sed, crenshaw2020sed}). Here, we employ the template perturbation algorithm introduced in \citealt{crenshaw2020sed} (hereafter CC20) and demonstrate that it significantly improves \ptz~accuracy for the CSST mock catalog, even with a very small training sample.

In Section~\ref{sect:data}, we briefly introduce the CSST mock catalog and the initial SED templates used for calibration. Section~\ref{sect:method} describes the perturbation algorithm from CC20. Details of the calibration process are presented in Section~\ref{sect:training}. Section~\ref{sect:results} compares the \ptz~results obtained using the SED templates before and after calibration. Finally, our conclusions are summarized in Section~\ref{sect:summary}.

\section{Data}
\label{sect:data}

\begin{figure}
    \centering
    \includegraphics[width=0.48335\textwidth]{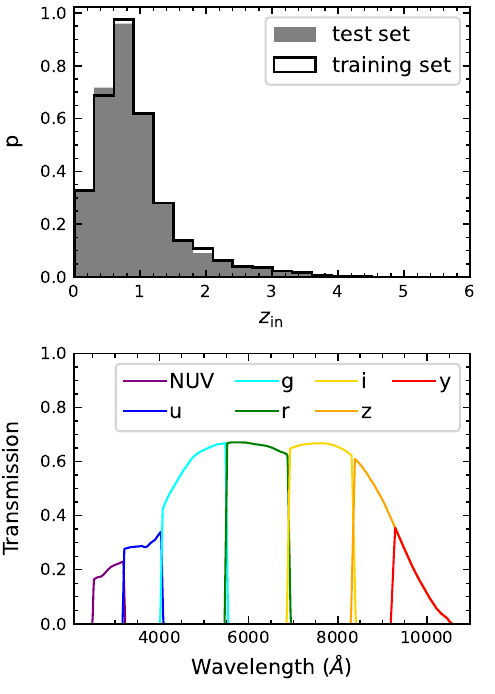}
    \caption{Top: Normalised redshift distribution of training set and test set from CSST mock catalog. Bottom: Response function of filters at NUV, $u, g, r, i, z$ and $y$ band for CSST. }
    \label{fig:zdist_filt}
\end{figure}

A CSST mock catalog was employed to calibrate the SED templates, and evaluating the improvement in \ptz~performance achieved by applying calibrated SED templates within the SED fitting algorithm. This catalog has also been used in previous CSST \ptz~assessment studies (\citealt{Zhou2021, Zhou2022, Lu2024rf, luo2024imputation, luo2024photometric}), which explored both machine learning and SED fitting methods.

In contrast to traditional approaches that rely on the mock catalogs generated from the light cone of dark matter N-body simulations, our mock catalog is derived from CSST like galaxy image simulation.
The simulation process is summarized below, more detailed information can be found in \citealt{Zhou2022} and \citealt{Lu2024rf}. First, HST/F814W images from COSMOS were rescaled to match the CSST pixel size, and square stamped images for each galaxy were created. The size of stamps is based on the length of galaxies' semi-major axis. Second, 31 SED templates from \textit{LePhare} \citealt{Arnouts1999_, ilbert2006calibrate} were assigned to each galaxy by fitting the photometric data at the given photo-$z$s from the COSMOS2015 catalog \citealt{laigle2016cosmos2015}. The theoretical flux calculated by convolving the response function of CSST filters with SED templates was taken as the simulated flux for each galaxy. Finally, simulated CSST like images were generated in these 7 bands based on the simulated flux, taking into account the exposure time, aperture size, and background noise level of CSST. Forced photometry was then applied to the simulated images to produce the mock catalog used in this work. Fig. \ref{fig:zdist_filt} shows the response function of each filter, with a wavelength coverage from $\sim2,500$ to $\sim11,000$\,\(\text{\AA}\).

\begin{figure}
    \centering
    \includegraphics[width=0.48335\textwidth]{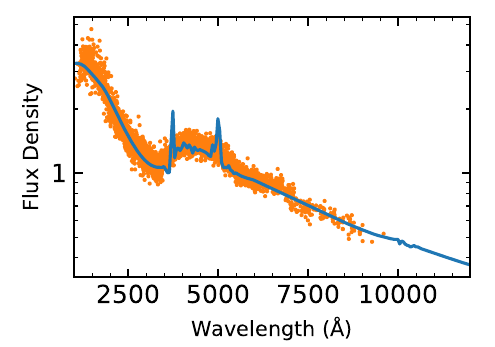}
    \caption{The photometric data of all the galaxies matched to the example SED template. The SED template has been normalized to 5500 \(\text{\AA}\), y-axis is in logarithmic scale.}
    \label{fig:example_matching}
\end{figure}

\section{The Perturbation Algorithm}
\label{sect:method}

In this section, we provide a brief introduction to the perturbation algorithm described in CC20 and \cite{Budavari2000sed}. The basic idea of the perturbation algorithm is to calibrate the SED templates according to the discrepancies between the photometric data and the SED templates. When redshifts are already known, the photometric data can be assigned to the SED templates with the most similar colors. Figure~\ref{fig:example_matching} shows an example in which all photometric data points are overlaid with their best match SED template in the rest frame. With a sufficiently large number of galaxies in the catalog, the overlapping region of their photometric data points can effectively reconstruct the SED's continuum. As the sample expands, it can even recover higher resolution features of the SED to a considerable degree.

However, discrepancies may exist between the distribution of photometric data and the SED templates. To address this, we calibrate the SED templates by applying perturbations, enabling better alignment between the templates and the photometric data. The templates perturbation is done in the rest frame. The flux densities derived by SED templates are
\begin{equation}
    F_n^\mathrm{temp}=\sum_k s_{k}{r\prime}_{n,k},
    \label{eq:F_temp_dispersed}
\end{equation}
where $s_k$ and ${r'}_{n,k}$ represent the flux density of the SED templates and the normalized filter response function in discrete wavelength bins in the rest frame, respectively.
 
All galaxies in the training set will be matched to the SED template with the most similar color before perturbation. The galaxies are separated into different groups referred to as the sub-training sets of each SED template. The cost function is in the form of
\begin{equation}
    \chi^2=\sum_{i}\sum_{n}\frac{{({\hat{F}}_{i,n}^\mathrm{temp}-F^\mathrm{obs}_{i,n})}^2}{{\sigma_{i,n}}^2}+\sum_{k}\frac{{({\hat{s}}_k-s_k)}^2}{{\Delta_k}^2}, 
    \label{eq:cost_func}
\end{equation}
where $F^{\mathrm{obs}}_{i,n}$ is the normalized observed flux density (see Sect. \ref{subsect:training_set}), and ${\hat{F}}_{i,n}^\mathrm{temp}$ represents the flux density derived from the perturbed SED templates, which is denoted as ${\hat{s}}_k$. The subscript $i,n$ indicates band $n$ of the $i$th galaxy in the sub-training set. The first term of the cost function represents how well the perturbed SED fits its sub-training set, weighted by the fractional error of observed flux densities $\sigma_i,n$. The second term is a penalty term, weighted by the parameter $\Delta_k$, which constrains the extent of the shape change during template perturbation and thus helps stabilize the shape of the perturbed SED. By minimizing this cost function, the shape change of the SED template, defined as $\xi_k = \hat{s}_k - s_k$, can be determined. 

\begin{figure}
    \centering
    \includegraphics[width=0.48335\textwidth]{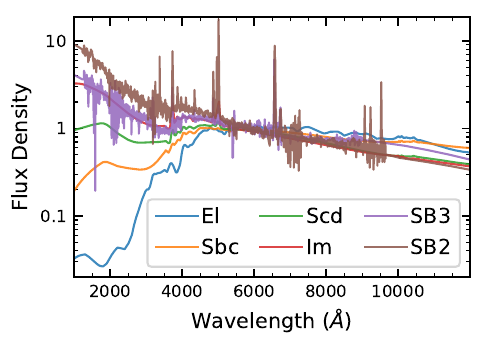}
    \caption{Initial SED templates of CWW+KIN. All templates are normalized at 5500 \(\text{\AA}\).}
    \label{fig:init_sed}
\end{figure}

For the part of hyperparameter optimization, CC20 implemented a uniform metric $\mathrm{\Delta}$ to replace $\mathrm{\Delta_k}$ in various templates and introduced an innovative parameter $w$, defined as the ratio of the loss function to the penalty term, thus regulating the pace of training.
\begin{equation}
    w=\frac{\sum_{k}{\Delta^{-2}{({\hat{s}}_k-s_k)}^2}}{\sum_{i}\sum_{n}{{\sigma_{i,n}}^{-2}({\hat{F}}_n^\mathrm{temp}-F^\mathrm{obs}_{i,n})}^2}, 
\end{equation}
by assuming
\begin{equation}
    \frac{\sum_{k}{({\hat{s}}_k-s_k)}^2}{\sum_{i}\sum_{n}{({\hat{F}}_n^\mathrm{temp}-F^\mathrm{obs}_{i,n})}^2}\sim\frac{N_k}{N}, 
\end{equation}
where $N$ is the number of photometric data points and $N_k$ is the number of wavelength bins. Then, $\Delta$ can be approximated as
\begin{equation}
    \Delta\approx\bar{\sigma}\sqrt{\frac{N_k}{wN}},
\end{equation}
where $\bar{\sigma}=\sum_{i}\sum_{n}\sigma_{i,n}$. 

Since the profile of SED templates changes after every perturbation, the photometric data matching results may differ from the previous perturbation. The algorithm will be iterated multiple times to obtain the best matching between the photometric data and the SED templates. Following the methodology outlined in CC20, we use the weighted mean squared error ($w$MSE) between the flux densities derived from the templates and the observed flux densities to quantify the alignment between the sub-training sets and the SED templates.
\begin{equation}
    w{\mathrm{MSE}}=\frac{1}{N}\sum_{i}\sum_{n}{\frac{1}{{\sigma_{i,n}}^2}{(F_{i,n}^\mathrm{temp}-F^\mathrm{obs}_{i,n})}^2}.
\end{equation} 
The termination of the perturbation process is determined by the relative change in $w$MSE, denoted as $d$MSE:
\begin{equation}
    d{\mathrm{MSE}}\ =\left|\ \frac{{w{\mathrm{MSE}}}_0-w{\mathrm{MSE}}}{{w{\mathrm{MSE}}}_0}\right|,
\end{equation}
The $w$MSE is calculated each time the SED template is perturbed or a new sub-training set is created. When $d$MSE falls below a specific threshold (4\% in this work) or reaches the maximum number of perturbations, we consider the SED templates to be sufficiently calibrated for the current iteration. The calibration process will stop when it reaches the maximum number of iterations or no templates are perturbed during the current iteration. For more details on the perturbation algorithm, we refer the reader to the studies of CC20 and \citealt{Budavari2000sed}

\section{SED Template Calibration}
\label{sect:training}

In this section, we describe in detail the process of calibrating the SED templates using the perturbation algorithm, including the creation of training sets, smoothing of initial templates to obtain more stable results, and comparisons between the calibrated and initial SED templates.

\subsection{Training set and initial SED templates}
\label{subsect:training_set}

To improve the precision of \ptz~estimation, we applied a signal-to-noise ratio (SNR) cut with a threshold of $\geq 10$ in either the $g$ or $i$ bands for both training and test sets. This criterion is consistent with the sample selection used in previous \ptz~evaluation studies for the CSST project (\citealt{Zhou2021, Zhou2022, Lu2024rf, luo2024imputation, luo2024photometric}). The final sample consists of 44,991 galaxies.

Instead of training the SED templates from scratch, we calibrate existing SED templates, enabling the use of a relatively small training set. However, utilizing a larger training set when available generally will result in a more stable outcome. In this work, we selected 10\% of the galaxies as the training set, which is $\sim$ 4,500 galaxies with a maximum redshift of about 4.83, while the remaining 90\% galaxies were the test set. The impact of the size of the training set will be discussed in Sect. \ref{sect:results}. Fig. \ref{fig:zdist_filt} illustrates the redshift distribution for both training and test sets, which represents an ideal scenario where the redshift distributions of the two sets are identical.

Perturbation will be applied to the SED templates based on their sub-training sets. For each galaxy, the photometric data are then matched to the SED template with the smallest $\chi^2$ value, representing the closest color match.
\begin{equation}
    \chi^2=\sum_{n}{\frac{1}{{\delta F}_n^\mathrm{obs}}{(F_n^\mathrm{temp}-F_n^\mathrm{obs})}^2},
    \label{eq:photomatch}
\end{equation}
where $\delta F_n^\mathrm{temp}$ is the uncertainty of normalized flux densities.

Due to variations in the intrinsic luminosity and distance of galaxies, the observed photometric data $f_n^\mathrm{obs}$ and its uncertainty $\delta f_n^\mathrm{obs}$ must be normalized before matching with the SED templates. Instead of using the median of $F^{\mathrm{temp}} / f^{\mathrm{obs}}$ as the normalization coefficient which is adopted in CC20, we apply the normalization method from the EAZY code, which is formulated as
\begin{equation}
    c_\mathrm{norm}=\frac{\sum_{n}{F_n^\mathrm{temp}f_n^\mathrm{obs}/({\delta f}_n^\mathrm{obs})^2}}{\sum_{n}{(F^\mathrm{temp}_n)^2/({\delta f}_n^\mathrm{obs})^2}},
    \label{eq:ez_norm}
\end{equation}
where the uncertainties of observed flux density are considered.

Since the template perturbation process relies on sub-training sets of each SED template, different normalization methods result in varied sub-training sets, which in turn lead to variations in the training results. An accurate matching between photometric data and SED templates is crucial for creating well-suited sub-training sets for each template, thereby enhancing the accuracy of photometric redshift (\ptz) estimation. The impacts of using different normalization coefficients will be discussed in Section \ref{sect:results}.

In principle, the initial templates can be naive log normal curves (CC20) or any set of SED templates. The initial templates should be comprehensive enough to cover a variety of galaxy types. Simultaneously, the number of templates should be minimized to ensure that each template has sufficient photometric data during the calibration process. Tab. \ref{tab:photo-zs} shows the results of \ptz~derived from different SED templates\footnote{Before calibrating the SED templates.}, where the fraction of outliers $\eta$ and $\sigma_{\mathrm{nMAD}}$ are defined in Sect. \ref{sect:results}. In this work, the CWW + KIN (\citealt{Coleman1980, Kinney1996, benitez2004ck}) template set is chosen as the initial SED templates (Fig. \ref{fig:init_sed}), due to its good photo-$z$ results and a relatively uniform distribution of the number of galaxies across each sub-training set. 

\begin{table}
    \centering
    \caption{Photo-$z$ results for different SED template set by using EAZY single template fitting mode.}
    \label{tab:photo-zs}
    \begin{tabular}{|l|c|c|}
    \hline
    SED Template & $\eta$ & $\sigma_{\mathrm{nMAD}}$\\
    \hline
    CWW+KIN                   & 3.85\%     & 0.047\\
    tweak\_fsps\_QSF\_12\_v3  & 7.47\%     & 0.052 \\
    COSMOS31                  & 6.46\%     & 0.044 \\
    EAZY\_V1.2\_dusty         & 17.25\%    & 0.073 \\
    \hline
    \end{tabular}
\end{table}

\subsection{Smoothing SED templates}
\label{subsect:sed_smoothing}

\begin{figure}
    \centering
    \includegraphics[width=0.48335\textwidth]{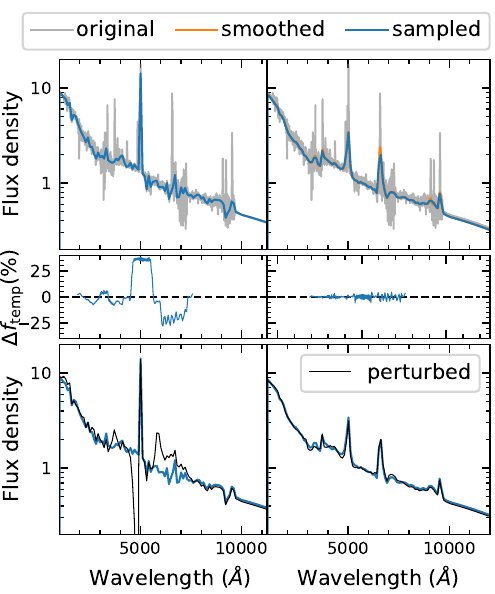}
    \caption{Impacts of sampling the SED templates. The left and right panels display the SED template before and after the smoothing process, respectively. 
    The upper panels show the SED template before and after sampling, represented by the gray and blue curves, with the smoothed template shown in orange in the upper right panel. 
    The middle panels show the relative error of the $F^\mathrm{temp}_i$ between the initial and sampled SED template at different redshifts. Note that the wavelength here indicate the mean wavelength of i band at different redshifts. 
    The lower panels illustrate the sampled SED template before and after perturbation, shown in blue and black curves, respectively. 
    All panels share the same $x$-axis, and the horizontal panels share the same $y$-axis.}
    \label{fig:temp_smooth}
\end{figure}

\begin{figure}
    \centering
    \includegraphics[width=0.48335\textwidth]{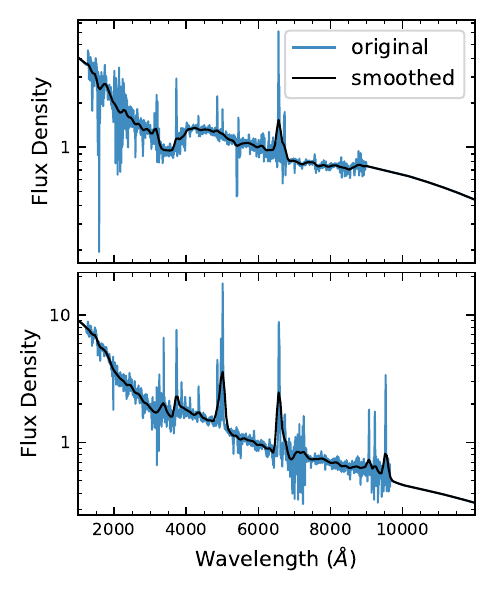}
    \caption{Two starburst templates before and after smoothing: SB3\_B2004a (top) and SB2\_B2004a (bottom). The blue and black curves represent the original and smoothed templates, respectively, with all templates normalized at 5500 \(\text{\AA}\).}
    \label{fig:smoothed_temp}
\end{figure}

A high resolution SED template with dense sampling on the wavelength grid will significantly increase the computational time required for the perturbation algorithm. To simplify the computation, the initial SED template is sampled using a fixed interval wavelength grid. However, for those SED templates with many high resolution features, such as strong emission or absorption lines, this sampling process can greatly impact $F^\mathrm{temp}$.

Fig. \ref{fig:temp_smooth} illustrates these effects using the SB2\_B2004a template (hereafter SB2) as an example, showing the differences in the SED and $F^\mathrm{temp}_i$ between the original template and a smoothed version. Both templates are sampled with a fixed wavelength step of 100~\text{\AA}. At 5000~\text{\AA}, where the peak of the $OIII$ emission line aligns with the sampling grid, the flux density derived from the $i$ band ($F^{\mathrm{temp}}_{i}$) can increase by up to 30\%. In addition, at 6500 \(\text{\AA}\), where the $H_\alpha$ emission line is missed due to sampling, the $F^\mathrm{temp}_{i}$ is 25\% lower than actual. 

These discrepancies in $F^\mathrm{temp}$ can lead to overcorrection in the perturbed SED template, as shown in the lower left panel of Fig. \ref{fig:temp_smooth}. Although the shape changes in the perturbed SED template effectively compensate for the discrepancies shown in the middle left panel, the presence of negative flux values (at 5,000 \(\text{\AA}\)) and an exaggerated, unrealistic absorption feature render the SED template unusable.

To avoid overcorrection, one approach is to modify the cost function by introducing an additional penalty term that further constrains changes in the SED shape, as suggested in \cite[ZEBRA]{feldmann2006zebra}. Increasing the sampling grid density or smooth the templates before calibration or smooth the templates before perturbation can also effectively reduce the overcorrection. In this work, we choose to smooth the SB2 template using a one dimensional Gaussian filter with a $\sigma$ of 100~\text{\AA}. 

As shown in the right panels of Fig. \ref{fig:temp_smooth}, the flux densities of the smoothed SED templates change smoothly near emission lines. The difference in $F^\mathrm{temp}_{i}$ between the sampled and smoothed initial templates is significantly reduced, and no overcorrection is observed in the perturbed SB2 template. Two starburst templates, SB3\_B2004a and SB2\_B2004a, due to their strong emission lines and other dense small features, are smoothed before calibration (Fig. \ref{fig:smoothed_temp}).

\subsection{Calibrating SED templates}
\label{subsect:temp_tr}
\begin{figure*}
    \centering
    \includegraphics[width=1\textwidth]{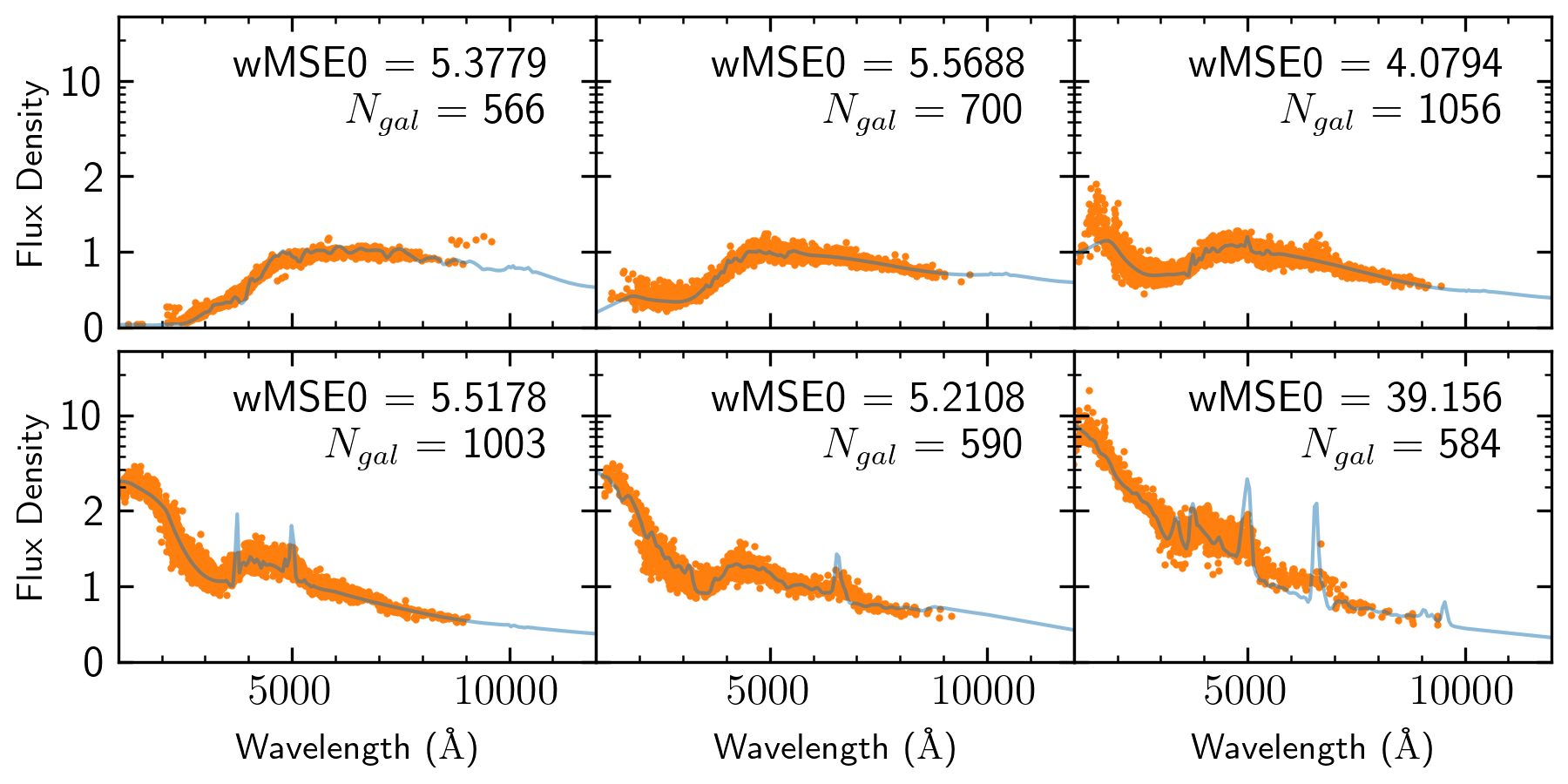}
    \caption{Initial SED templates with their sub-training sets for the first iteration. The blue curves are SED templates; orange dots are the photometric data. The flux density of all SED templates is normalized to 1 at 5500 \(\text{\AA}\). Note that we use a linear scale for the y-axis below 2, while maintaining a logarithmic scale for values above 2 in this figure.}
    \label{fig:untr_sub_tr_sets}
\end{figure*}

\begin{figure*}
    \centering
    \includegraphics[width=1\textwidth]{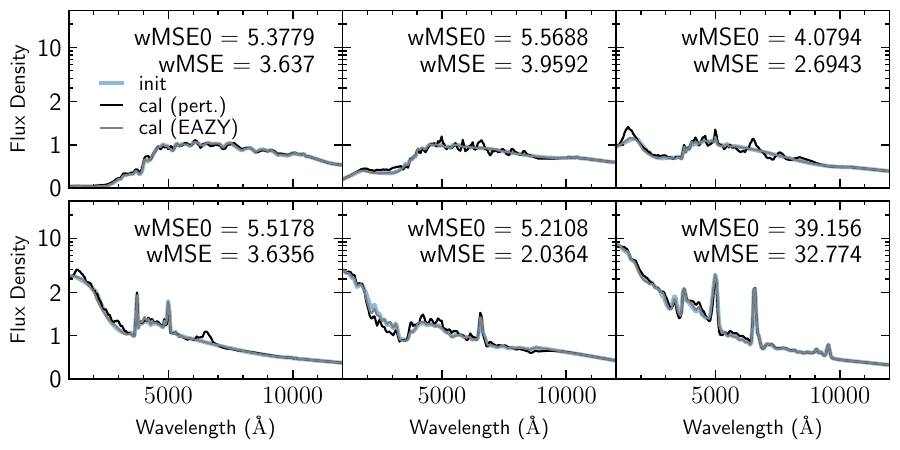}
    \caption{Calibrated templates and initial templates. The blue curves represent the initial templates, black and grey curves are the templates calibrated with perturbation algorithm and EAZY. Only initial templates are normalised at 5500 \(\text{\AA}\). To make the changes in the SED profile more apparent, we use a linear scale for the y-axis below 2, while maintaining a logarithmic scale for values above 2.}
    \label{fig:tr_untr_temp}
\end{figure*}

\begin{figure}
    \centering
    \includegraphics[width=0.48335\textwidth]{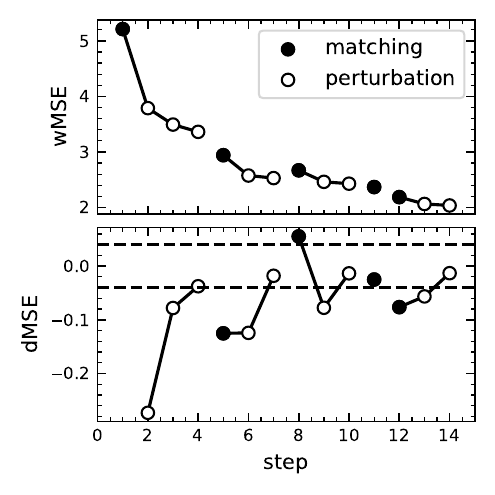}
    \caption{An example of the perturbation process for the SB3\_B2004a template. The upper and lower panels show the $w$MSE and $d$MSE throughout the perturbation process, respectively. Open circles represent the $w$MSE and $d$MSE after each perturbation, solid markers represent the $w$MSE and $d$MSE values after the creation of new sub-training sets. Changes in MSE and $d$MSE during the same iteration are connected by solid lines. The dashed lines in the lower panel denote the $\pm$4\% threshold for $d$MSE.}
    \label{fig:wMSE_dMSE_compare}
\end{figure}

We utilized the template perturbation algorithm described in Sect. \ref{sect:method} to calibrate the CWW+KIN templates, employing photometric data sourced from the CSST mock catalog as detailed in Sect. \ref{sect:data}. 
According to CC20, the training results are robust to $w$, the CWW + KIN templates are calibrated with $w=1.0$, which is similar with the $w$ used in the work of CC20. Given that the final few iterations have minimal impact on the SED templates (as shown in Fig. \ref{fig:wMSE_dMSE_compare}), we set a maximum of 3 perturbations per iteration and a maximum of 5 iterations for the calibration process. The threshold $d$MSE is set to 4\%. In Fig. \ref{fig:untr_sub_tr_sets}, we present the initial SED templates together with their corresponding sub-training sets. The distribution of galaxies among the sub-training sets appears relatively uniform for each SED template,  ranging from approximately 1000 galaxies to at least 500 galaxies per template. Most templates show similar initial $w$MSE, approximately 5. However, the initial $w$MSE of the bluest template, SB2, is approximately one order of magnitude larger than the others. This difference arises from the photometric data matching process, during which all templates are normalized to a wavelength of 5500 \(\text{\AA}\). The higher $w$MSE of SB2 is attributed to its larger template flux density at the blue end compared to the other templates.

After five iterative processes, the calibrated SED templates and the initial SED templates are shown in Fig. \ref{fig:tr_untr_temp}. The changes in the profile of the SED templates are not obvious, in order to make the changes in the SED profile more apparent, we use a linear scale for the y-axis below 2, while maintaining a logarithmic scale for values above 2. Each template shows a decrease in $w$MSE to varying degrees after calibration, suggesting a more accurate alignment with the photometric data, when comparing with the original templates. 

The SB3 template is illustrated as an example (Fig. \ref{fig:wMSE_dMSE_compare}) to demonstrate the evolution of $w$MSE and $d$MSE throughout the calibration process. The initial stages show a rapid decline in $w$MSE, indicating that the first few perturbations are usually the most influential. Which suggest it's crucial to achieve a precise alignment between the photometric data and the SED templates during this stage. After five iterations, the final $w$MSE of the calibrated SB3 template experiences a substantial reduction of over 60\%, dropping from 5.211 to 2.036. This demonstrates the efficacy of the calibration process in enhancing the alignment between the template and the observed photometric data. 

\begin{table}
    \centering
    \caption{Flux offsets used for absolute calibration given by EAZY, i band is taken as the reference band.}
    \label{tab:zero-point}
    \begin{tabular}{|l|c|c|c|c|}
    \hline
    band & NUV & u & g & r\\
    \hline
    flux offset & 0.9229 & 0.9730 & 0.9683 & 0.9695\\
    \hline
    band & i & z & y & \\
    \hline
    flux offset & 1.0000 & 1.0376 & 1.0549 & \\
    \hline
    \end{tabular}
\end{table}

In addition to the perturbation algorithm proposed in CC20, the SED fitting code EAZY (\citealt{Brammer2008}) can also iteratively calibrate the SED templates according to the distribution of photometric data, a process known as Absolute Calibration (\citealt{weaver_cosmos2020_2022}). Additionally, the flux of each band can also be corrected with a offset. The calibration process is similar to the perturbation method: the photometric data is first matched with the SED templates, and then the median value of $F^\mathrm{obs}_n / F^\mathrm{temp}_n$ in each rest-frame wavelength bin is calculated, as well the offsets for each band. The SED templates are calibrated by multiplying the calculated ratio with the flux density of the SEDs (\citealt{ilbert2006calibrate}), the calibration process will be iterated until the flux offsets in each band changed by less than 4\%. EAZY can utilize either spec-$z$ or photo-$z$ for flux and SED calibration.

In this work, we used the same training set as in the perturbation algorithm, where all galaxies have available spec-$z$ values and parameters, to calibrate the SED templates using EAZY. The process converged after 4 iterations, and the final flux offsets for the different bands are shown in Tab. \ref{tab:zero-point}. The calibrated SEDs are displayed in Fig. \ref{fig:tr_untr_temp} as grey curves.

While the calibrated SED templates from EAZY exhibit the same general trend as those calibrated with the perturbation algorithm, the amplitude of the changes in the SED profile is less apparent. This is because the calibration method in EAZY splits the wavelength into larger bins when using a small training set, resulting in smaller median values for the residuals and, consequently, smaller changes in the SED profiles. The changes in the SED profiles would be more apparent when using the full set of galaxies ($\sim$ 45,000) as the training set, which further suggests that misalignment between the photometric data and the SED templates may occur on smaller wavelength scales. In contrast, the perturbation algorithm effectively corrects misalignment on a small wavelength scale, as it directly adjusts the SED templates along the wavelength grid without binning. This precise adjustment is particularly efficient when the discrepancies between $F^\mathrm{temp}$ and $F^\mathrm{obs}$ are not large.

\section{Results}
\label{sect:results}

We used the \ptz~results of galaxies in the test set, which were derived from the calibrated templates, to quantify improvements in the SED templates after calibration. This section details the \ptz~results estimated with the SED fitting code EAZY, and discusses the impact of different setups during the calibration process.

\subsection{Photo-z estimation}
The SED fitting code EAZY determines redshifts by maximizing the redshift probability distribution function (PDF) which in format of
\begin{equation}
    p(z)=e^{-0.5(\chi_z^2-\chi_\mathrm{min}^2)},
\end{equation}
where the $\chi_z^2$ is derived from
\begin{equation}
    \chi_{z,i}^2=\sum_{n=1}^{N_\mathrm{filt}}\frac{{(F^\mathrm{temp}_{z,i,n}-F^\mathrm{obs}_n)}^2}{{(\delta F^\mathrm{obs}_n)}^2}.
\end{equation}

\begin{figure}
    \centering
    \includegraphics[width=0.48335\textwidth]{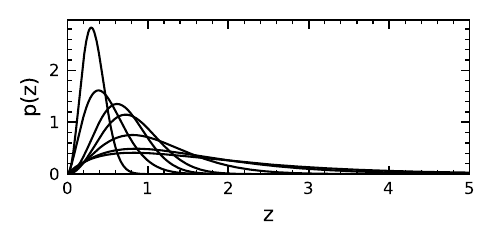}
    \caption{Prior probabilities for different $i$-band magnitudes, ranging from 20.5 (narrowest) to 26.5 (widest).}
    \label{fig:prior}
\end{figure}

Photo-$z$s may become catastrophic outliers due to degeneracies in the color-redshift space, an incomplete set of templates, or poor photometric data (\citealt{Newman2022}). Despite a slight decrease in accuracy at high redshifts, a straightforward and effective method to overcome these degeneracies is to apply a Bayesian prior based on the redshift distribution. \cite{Benitez2000} pioneered this approach in BPZ, and EAZY implements a similar prior to constrain the redshift probability distribution function (PDF), as described in \cite[Eq. 3]{Brammer2008}:
\begin{equation}
    p(z|m_{0,i})\propto z^{\gamma_i}{\rm {exp}}[-\left(z/z_{0,i}\right)^{\gamma_i}],
\end{equation}
where $\gamma_i$ and $z_{0,i}$ represent the parameters, the probability density function $p(z)$ is dependent on magnitude $m_{0,i}$ of the reference band. The PDF of galaxies can be calibrated by multiplying with prior probability density directly. 
Fig. \ref{fig:prior} shows the part of the Bayesian prior used in this work, which is based on the $i$ band photometric data and the distribution of $z_\mathrm{in}$. 

For each galaxy, we use $z_\mathrm{peak}$ as its best photo-$z$, which is estimated by the mean value of the largest peak in PDF, where $z_\mathrm{peak}=(\int z p\left(z\right)dz)/(\int p\left(z\right)dz)$.
We employ the normalized median absolute deviation $\sigma_\mathrm{nMAD}$, redshift bias $b$, and the fraction of outliers $\eta$ to quantify the accuracy and precision of photo-$z$ estimations, defined as follows:
\begin{equation}
    \sigma_\mathrm{nMAD}=1.48\times \mathrm{median}(|\frac{\Delta z-\mathrm{median}(\Delta z)}{1+z_\mathrm{in}}|), 
\end{equation}
\begin{equation}
    b=\langle dz \rangle, 
\end{equation}
\begin{equation}
    dz=\frac{z_\mathrm{peak}-z_\mathrm{in}}{1+z_\mathrm{in}}.
\end{equation}
Outliers are defined as galaxies with $|dz|>0.15$. In this work, all photo-$z$ s are derived using single template fitting mode, with a Bayesian prior based on the $i$ band. The amplitude of the template error function is set to 0, and the other parameters remain EAZY's default.

\begin{figure*}
    \centering
    \includegraphics[width=\textwidth]{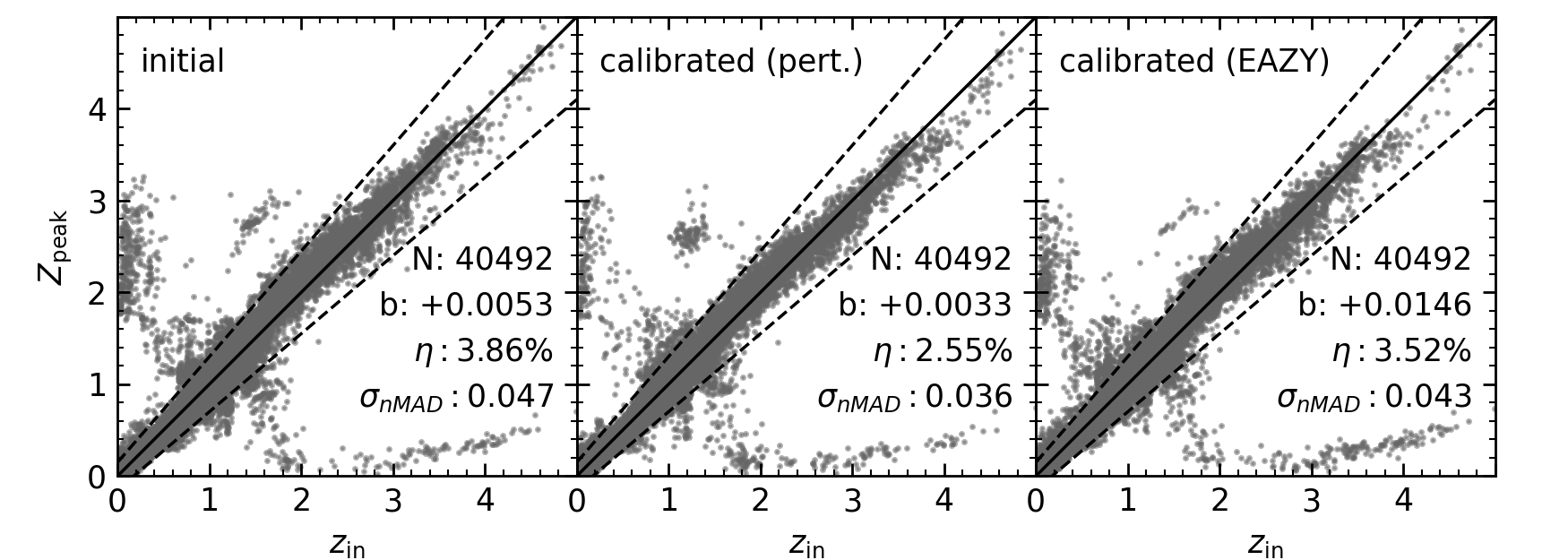}
    \caption{Photo-$z$ results of 40492 galaxies in test set. Each panel shows the $z_\mathrm{peak}$ vs $z_\mathrm{in}$ from initial CWW+KIN templates (left), CWW+KIN templates calibrated with perturbation algorithm (middle) and CWW+KIN templates calibrated with EAZY (right), respectively. The dash lines and solid line indicate $dz=0.15$ and $z_\mathrm{in}=z_\mathrm{peak}$ respectively.  }
    \label{fig:photo-zs}
\end{figure*}

\begin{figure}
    \centering
    \includegraphics[width=0.48335\textwidth]{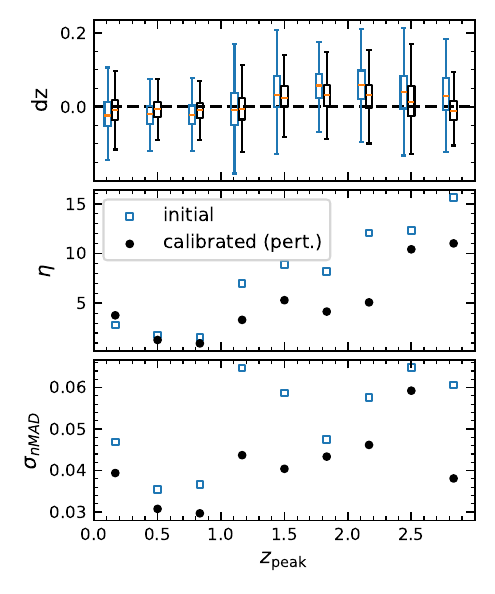}
    \caption{The accuracy parameters of different photo-$z$  bins. From top to bottom are dz, $\eta$ and $\sigma_\mathrm{nMAD}$ respectively. Results of calibrated and initial template are shown in solid dots and open squares. dz is shown in boxplot, where the orange bar indicates the median value, box indicate the quartile and the error bar is derived with 1.5 times the inter-quartile range.}
    \label{fig:photo-z_zbin}
\end{figure}

\begin{figure}
    \centering
    \includegraphics[width=0.48335\textwidth]{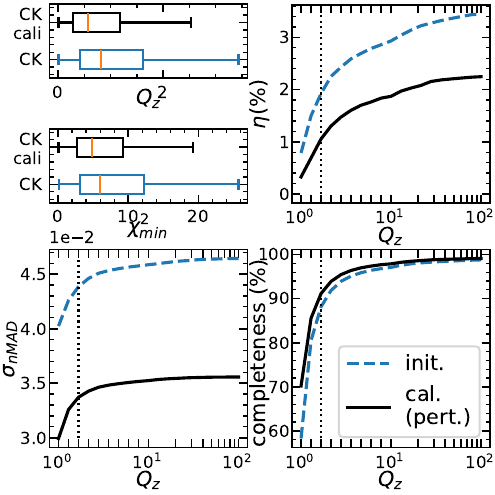}
    \caption{The distribution of $Q_z$ and $\chi^2_\mathrm{min}$ for two template sets and the photo-$z$ results with different $Q_z$ threshold. The upper left pannels are the boxplots of the distribution of $Q_z$ and $\chi^2_\mathrm{min}$. The photo-$z$ results before and after calibration are shown as blue dash and black solid curves respectively. The orange bars are the median value of the distribution. The dashed line indicate the $Q_z$ = 3 cut we applied.}
    \label{fig:qz_cut}
\end{figure}

\subsection{Photo-z results}

The photo-$z$ results for the test set are shown in Fig. \ref{fig:photo-zs}, using the template calibrated with the perturbation algorithm. The corresponding values for $\eta$ and $\sigma_\mathrm{nMAD}$ are 2.55\% and 0.036, respectively. Compared to the results obtained using the initial CWW + KIN template (3.86\% and 0.047), we observe reductions of approximately 30\% in $\eta$ and 20\% in $\sigma_\mathrm{nMAD}$. The general bias also decreased from +0.0053 to +0.0033. For templates calibrated with EAZY, which exhibit negligible changes in SED, yield \ptz~results similar to the initial ones, with values $\eta$ and $\sigma_{\mathrm{nMAD}}$ of 3. 52\% and 0.043. In order to focus on the enhancements provided by the perturbation algorithm, and to simplify subsequent figures, we exclude the results from templates calibrated with the EAZY method.

The $z_\mathrm{peak}$ estimates for $z_\mathrm{in} > 3$ are slightly lower than the corresponding $z_\mathrm{in}$ values for both calibrated templates. This discrepancy is caused by the insufficient high redshift galaxies in our training set, which suggests when in a more realistic situation where the number of high redshift galaxies with spec-$z$ is limited, the photo-$z$ may be biased at higher redshift after template calibration.

In Fig. \ref{fig:photo-z_zbin}, we present d$z$, the fraction of outliers $\eta$, and $\sigma_\mathrm{nMAD}$ across different $z_\mathrm{peak}$ bins. Compared to the initial CWW + KIN template, the calibrated CWW + KIN template results in reduced or comparable $\eta$ and $\sigma_\mathrm{nMAD}$ for $z_\mathrm{peak} < 3$. The most significant improvements in $\eta$ and $\sigma_{\mathrm{nMAD}}$ occur in the range $1 < z_{\mathrm{peak}} < 2.2$, where the Balmer break has shifted out of the middle part of the CSST filters, and the Lyman break has not yet reached the middle part of the CSST filters. In this region, the continuum plays a crucial role in photo-$z$ estimation. The substantial improvements in the photo-$z$ results for $1 < z_\mathrm{peak} < 2.2$ are consistent with the results shown in Fig. \ref{fig:tr_untr_temp}, where the perturbation algorithm primarily modifies the continuum portion of the SEDs.

The application of prior eliminates a large number of outliers; however, there are still quite a few outliers in both \ptz s derived by calibrated and initial SED templates. Most of these outliers exhibit very low SNR in the NUV and / or $u$ bands. The lack of constraint of photometric data with high SNR makes these galaxies have multiple peaks in redshift PDFs, which makes their \ptz s less reliable.

A redshift quality exclusion criterion, based on Eazy's quality parameter $Q_z$ (\citealt{Brammer2008}, Eq. 8) is applied to remove possible outliers. $Q_z$ describes the reliability of photo-$z$ estimations, it is related with redshift PDF, $\chi^2$ and number of filters $N_\mathrm{filt}$ used in fitting, which in formula of
\begin{equation}
    Q_z=\frac{\chi^2_\mathrm{min}}{N_\mathrm{filt}-3}\frac{z_\mathrm{up}^{99}-z_\mathrm{lo}^{99}}{p_{\Delta z=0.2}},
\end{equation}
where $z_\mathrm{up}^{99}-z_\mathrm{lo}^{99}$ represent the 3$\sigma$ confidence intervals computed from the redshift PDF. $\chi^2_\mathrm{min}$ is the minimum chi-square value between the observed photometric data and the SED template, and $p_{\Delta z=0.2}$ is the probability of redshift in $z_\mathrm{peak} \pm 0.2$.

\begin{figure*}
    \centering
    \includegraphics[width=0.6667\textwidth]{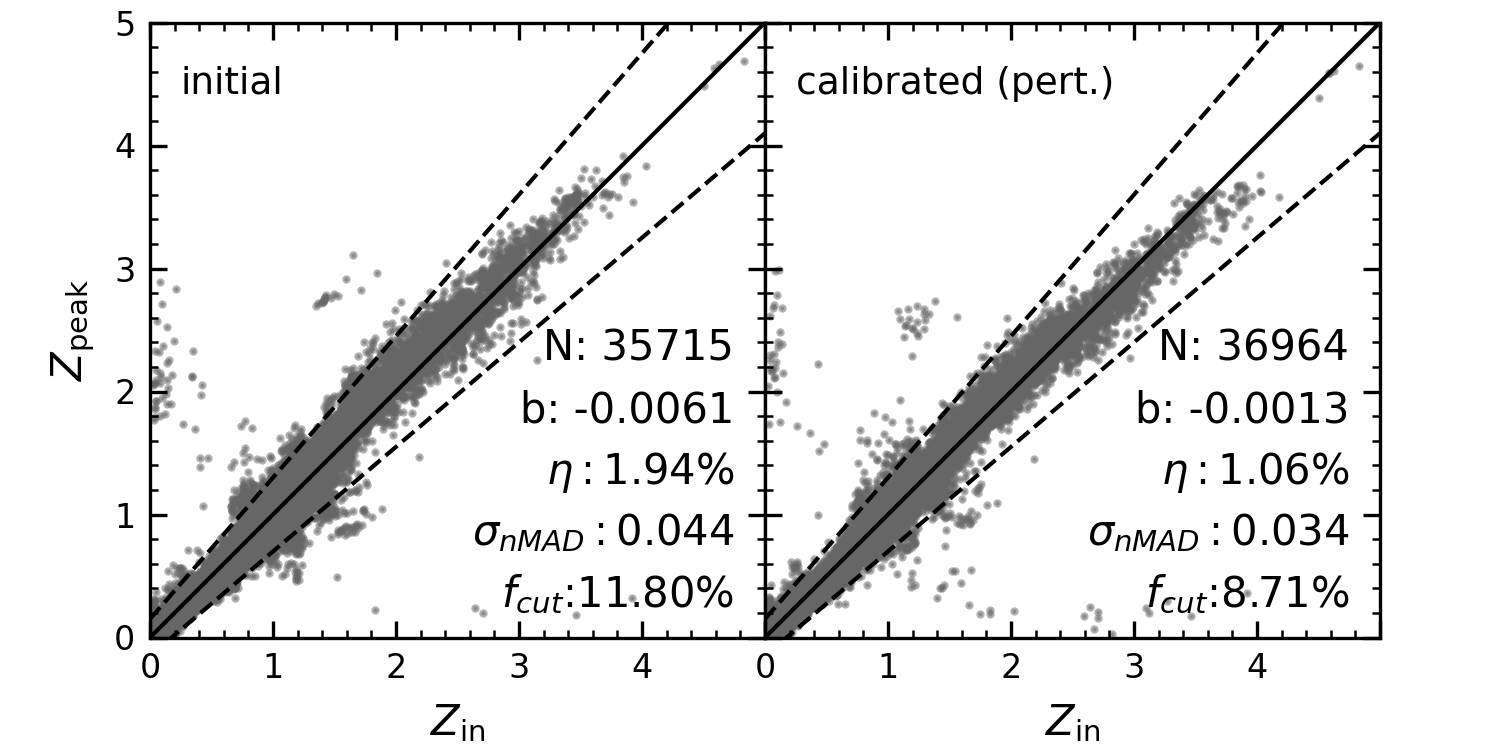}
    \caption{Results of \ptz~estimations after apply $Q_z$ cut.}
    \label{fig:photo-zs_cut}
\end{figure*}

\begin{figure}
    \centering
    \includegraphics[width=0.48335\textwidth]{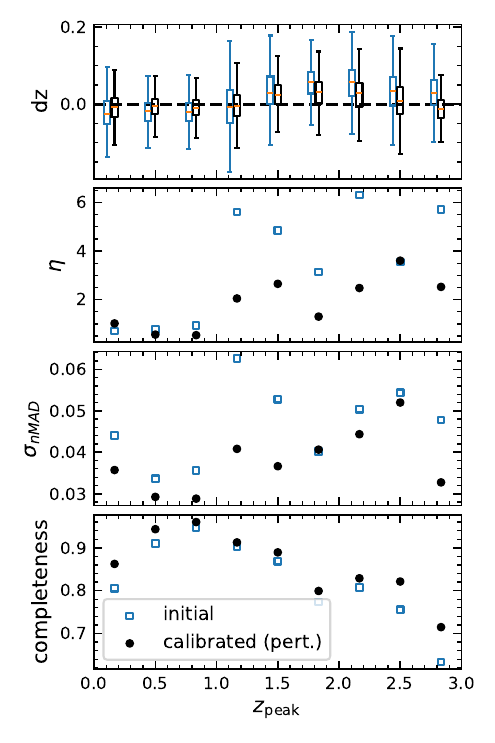}
    \caption{ Photo-$z$  results at different redshift bins after quality cuts. Panels from top to bottom are bias, fraction of outliers, $\sigma_\mathrm{nMAD}$ and completeness at different redshift bins, respectively.}
    \label{fig:photo-zs_cut_zbin}
\end{figure}

With the price of completeness, selecting based on $Q_z$ can further improve the photo-$z$ accuracy. Fig. \ref{fig:qz_cut} shows that the $Q_z$ and $\chi^2$ distribution of using the CWW + KIN template before calibration is much smaller than the values after calibration in the photo-$z$ estimation. This indicates that calibrated templates achieve a better match between photometric data and SEDs. This is consistent with previous results for the smaller $w\mathrm{MSE}$ shown in Fig. \ref{fig:wMSE_dMSE_compare}. A smaller $Q_z$ distribution is crucial for applying the redshift quality cut. Fig. \ref{fig:qz_cut} also displays the \ptz~results after applying various $Q_z$ thresholds. The calibrated templates consistently show improved performance across different $Q_z$ thresholds. We find that $Q_z=3$ is an effective threshold for maintaining lower $\eta$ and $\sigma_\mathrm{nMAD}$ while retaining a sufficient number of galaxies. However, the number of galaxies significantly decreases when $Q_z \leq 3$, as depicted in the lower right panel of Fig. \ref{fig:qz_cut}.

Fig. \ref{fig:photo-zs_cut} and \ref{fig:photo-zs_cut_zbin} show the photo-$z$ results after excluding galaxies with $Q_z > 3$. The fraction of galaxies excluded is reported as $f_\mathrm{cut}$. The fraction of outliers $\eta$ decreased significantly after quality cuts, for CWW + KIN templates, the $\eta$ and $\sigma_\mathrm{nMAD}$ are 1.94\% and 0.044 respectively, with a $f_\mathrm{cut}$ of 11.8\%. 
Compared to the initial templates, the result from calibrated CWW + KIN yield even lower $\eta$ and $\sigma_\mathrm{nMAD}$, which is 1.06\% and 0.034. Additionally, the calibrated template set retain slightly higher completeness after the $Q_z$ cut, with only 8.71\% of galaxies excluded. When compared to Fig. \ref{fig:photo-z_zbin}, Fig. \ref{fig:photo-zs_cut_zbin} shows a similar trend in the \ptz~results across different $z_\mathrm{peak}$ bins.

\begin{figure}
    \centering
    \includegraphics[width=0.48335\textwidth]{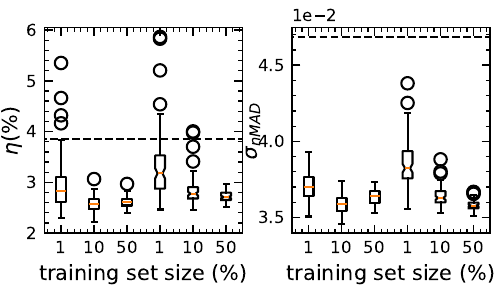}
    \caption{Boxplot represents photo-$z$ results across varying training set dimensions and different normalization method. Dashed lines denote the photo-$z$ results derived from the initial CWW+KIN template. The orange lines represent the median values of the photo-$z$ results obtained from the calibrated SED templates, while the boxes delineate the inter-quartile range. Error bars extend to 1.5 times the inter-quartile range, and circles identify outliers beyond these error bars. The results of normalizing with Eq. \ref{eq:ez_norm} and median of ratios are marked with box and box with notch, respectively.}
    \label{fig:photo-zs_tr_size}
\end{figure}

\begin{figure}
    \centering
    \includegraphics[width=0.48335\textwidth]{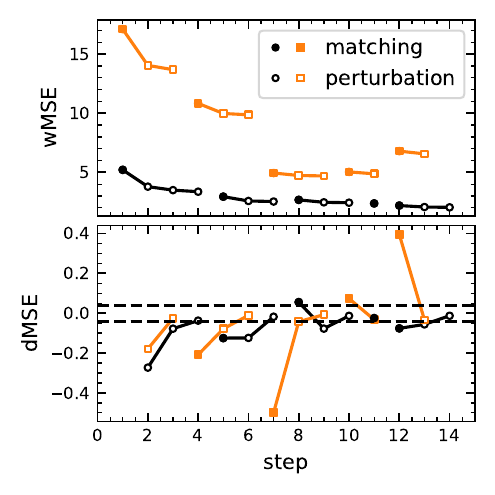}
    \caption{The variration of $w$MSE and $d$MSE during the perturbation process of SB3\_B2004a template. The upper and lower panel show the $w$MSE and $d$MSE through the perturbation process, respectively. Hollow markers represent the $w$MSE and $d$MSE after each perturbation, solid markers represent the $w$MSE and $d$MSE each time new sub-training sets are created, MSE and $d$MSE changes in each iteration are connected with solid lines. Dashed lines in the right panel indicate the 4\% threshold of $d$MSE. Normalizing with Eq. \ref{eq:ez_norm} and median of ratios methods are marked with black and orange colors, respectively.}
    \label{fig:mse_compare}
\end{figure}

\subsection{Impact of various setups}

To evaluate the impact of training set size on the results, we used training sets of three different sizes: 1\% ($\sim450$ galaxies), 10\% ($\sim 4 \times 10^3$ galaxies), and 50\% ($\sim 2 \times 10^4$ galaxies) of the catalog. For each training set size, samples were randomly selected from the catalog, and the templates were calibrated accordingly. This process was repeated 100 times to reduce potential biases introduced by sample selection. 
The results shown in Fig. \ref{fig:photo-zs_tr_size} indicate that a training set with $\sim450$ galaxies is sufficient to improve the photo-$z$ accuracy, especially in terms of $\sigma_\mathrm{nMAD}$. However, using such a small training set results in unstable improvements; for instance, 4 out of 100 tests exhibited a higher $\eta$ compared to the initial templates. Increasing the training set size can enhance the stability of the \ptz~results. For CWW+KIN template, we find that training sets with about 4,500 galaxies consistently yield improved \ptz~results.

Additionally, we tested the impact of using different normalization methods during the photometric data matching process. All templates are calibrated with the same parameters. These results are also shown in Fig. \ref{fig:photo-zs_tr_size}. For SED templates calibrated with small size training sets (450 galaxies), normalizing with Eq. \ref{eq:ez_norm} yields better \ptz~results from calibrated templates. The \ptz~results become similar for both methods when the size of training set increased to $\sim 4,500$ galaxies. 

Fig. \ref{fig:mse_compare} shows an example of the $w$MSE and $d$MSE history during the calibration process using different normalization methods. The variations in the initial $w$MSE primarily result from the different definitions of two normalization coefficients. When normalizing with Eq \ref{eq:ez_norm}, $d$MSE caused by the matching process is much smaller compared to normalizing with median values or ratios. This indicates a more accurate alignment between the SED templates and photometric data within sub-training sets, which proves particularly beneficial when the training set size is small.

\begin{figure}
    \centering
    \includegraphics[width=0.48335\textwidth]{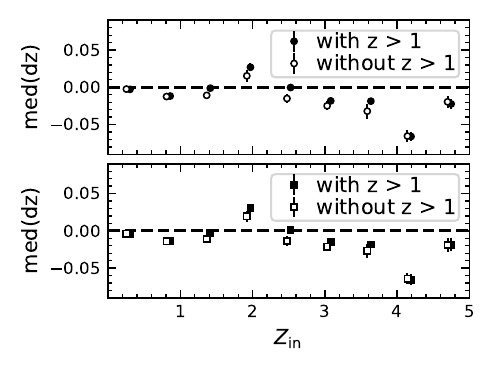}
    \caption{dz values from 100 different test sets, divided into different $z_{\mathrm{in}}$ bins. Solid and hollow circles (squares) represent the median values of the binned median dz values across the 100 results from SED templates calibrated with high redshift galaxies and without high redshift galaxies, respectively. Error bars denote the standard deviations of dz. The upper and lower panels show the results from training sets consisting of 10\% and 50\% of the SNR-cut catalog, respectively.}
    \label{fig:bias_test}
\end{figure}

As shown in Fig. \ref{fig:photo-zs}, the photo-$z$ results could be biased at high $z_{\mathrm{in}}$ due to the lack of high redshift samples in our training sets. To address this, we conducted an additional test by excluding galaxies with $z_{\mathrm{in}} > 1$ from the training sets. Fig. \ref{fig:bias_test} shows the dz values from 100 different test sets for different training set size, divided into different $z_{\mathrm{in}}$ bins. For galaxies with $z_{\mathrm{in}} > 4$, the photo-$z$ results are almost unaffected, which is expected since there are very few galaxies with redshifts greater than 4 in our training sets. However, for galaxies in the range $1 < z_{\mathrm{in}} < 4$, the templates calibrated without high redshift galaxies show slightly lower dz values, indicating that the estimated photo-$z$s become smaller after excluding galaxies with $z_{\mathrm{in}} > 1$ from the training sets, potentially leading to underestimated \ptz~at higher $z_{\mathrm{in}}$.

\section{Summary}
\label{sect:summary}
In this work, we used the perturbation algorithm described in CC20 to calibrate CWW+KIN templates for CSST mock catalog. The mock catalog is derived from simulated CSST images which based on HST-ACS F814W images. We selected galaxies with SNR $\geq$ 10 in g or i band as training set and test set for SED calibration.

Before the calibration, we smoothed the two SED templates of starburst galaxies with strong emission lines. The smoothed templates successfully avoided overcorrection and reduced the discrepancies in ${F}^\mathrm{temp}$ caused by sampling. We tested both the absolute calibration method used in EAZY and the perturbation algorithm for template calibration. Compared to absolute calibration, the perturbation algorithm can modify the SED templates directly without binning the wavelength, making it more sensitive to misalignments at small wavelength scale when the training set is small. For the CSST mock catalog, the SED templates calibrated with the perturbation algorithm yielded photo-$z$ results with higher accuracy.

We used SED fitting code EAZY for photo-$z$ estimation. The results derived from SED templates calibrated with the perturbation algorithm demonstrated moderate improvements: a $\sim30\%$ reduction in the fraction of outliers ($\eta$), reduced to 2.55\%, and a $\sim20\%$ reduction in the normalized median absolute deviation ($\sigma_{\mathrm{nMAD}}$), which decreased to 0.036. In contrast, templates calibrated using absolute calibration yielded photo-$z$ results with an $\eta$ of 3.52\% and a $\sigma_{\mathrm{nMAD}}$ of 0.043, similar to the initial CWW+KIN templates, which had $\eta$ of 3.86\% and $\sigma_{\mathrm{nMAD}}$ of 0.047. The calibrated SED templates provided a tighter fitting during the \ptz~estimation process, resulting in a improved redshift quality parameter ($Q_z$). Consequently, there was less loss in completeness after applying quality cuts of $Q_z\leq3$, the photo-$z$ results showed a $\sim$60\% decrease in $\eta$ and a similar $\sigma_\mathrm{nMAD}$, which is 1.06\% and 0.034, respectively. However, this quality cut still resulted in an exclusion of 8.71\% of the galaxies.

Additionally, we tested the impact of varying training set sizes and different normalization coefficients on photometric data matching process. We found that a training set of approximately 4,500 galaxies is sufficient to consistently improve the photo-$z$ results derived from calibrated templates, regardless of the normalization coefficient used. For smaller training sets (450 galaxies), employing the normalization coefficient defined by Eq. \ref{eq:ez_norm} yielded slightly better performance. 

The limited representation of high-redshift galaxies in our training set introduces a minor systematic underestimation of their photo-$z$s compared to the input redshift values, even in scenarios where the redshift distributions of both the training and test sets are identical, thereby establishing an ideal condition. In realistic observations, it is likely that fewer high redshift galaxies will have spec-$z$ available,  potentially introducing additional bias into the \ptz~results. The exclusion of galaxies with $z_\mathrm{in}>1$ exacerbates the \ptz~ underestimation for high redshift galaxies, relative to SED templates calibrated with $z_\mathrm{in}>1$ training data. Future extensions of this study will focus on incorporating hierarchical Bayesian priors in order to mitigate high redshift biases while preserving the method's computational efficiency and small-sample adaptability.

The template calibration approach, which utilizes a perturbation algorithm, exhibits the capacity to enhance accuracy with the deployment of a comparatively small training set. Moreover, it provides a feasible route for conducting initial cosmological analyses in stage-IV surveys such as CSST, where template calibration is capable of producing immediately applicable \ptz~ catalogs prior to the implementation of comprehensive spectroscopic campaigns.

\begin{acknowledgements}
WD acknowledges the science research grants from the China Manned Space Project with NO. CMS-CSST-2021-B01. ZJL acknowledges support from the Shanghai Science and Technology Foundation Fund under grant No. 20070502400, and the science research grants from the China Manned Space Project. LPF acknowledges the support from the Innovation Program of Shanghai Municipal Education Commission (Grant No. 2019-01- 07-00-02-E00032).  ZC acknowledges support from the China Manned Space Project with NO. CMS-CSST-2021-A07. YG acknowledges the support from National Key R\&D Program of China grant Nos. 2022YFF0503404,
2020SKA0110402, the CAS Project for Young Scientists in Basic Research (No. YSBR-092), and China Manned Space Project with Grant No. CMS- CSST-2021-B01. S.Z. acknowledges support from the National Natural Science Foundation of China (Grant No. NSFC-12173026), the Program for Professor of Special Appointment
(Eastern Scholar) at Shanghai Institutions of Higher Learning and the Shuguang Program of Shanghai Education Development Foundation and Shanghai Municipal Education Commission. ZF acknowledges the support from NSFC grant No. U1931210. This work is also supported by the National Natural Science Foundation of China under Grants Nos. 12141302 and 11933002, and the science research grants from China Manned Space Project with Grand No. CMS-CSST-2021-A01.

\end{acknowledgements}


\label{lastpage}

\bibliography{main}
\bibliographystyle{raa}

\end{document}